\documentclass[
    ,final            
  ]
  {aipproc}

\layoutstyle{6x9}

\begin{document}

\title{Optical photometric monitoring of gamma-ray binaries}


\author{Xavier Paredes-Fortuny}{
  address={Departament d'Astronomia i Meteorologia, Institut de Ci\`encies del
Cosmos (ICC), Universitat de Barcelona (IEEC-UB), Mart\'{\i} i Franqu\`es 1,
08028 Barcelona, Spain}
}

\author{Marc Rib\'o}{
  address={Departament d'Astronomia i Meteorologia, Institut de Ci\`encies del
Cosmos (ICC), Universitat de Barcelona (IEEC-UB), Mart\'{\i} i Franqu\`es 1,
08028 Barcelona, Spain}
}

\author{Octavi Fors}{
  address={Departament d'Astronomia i Meteorologia, Institut de Ci\`encies del
Cosmos (ICC), Universitat de Barcelona (IEEC-UB), Mart\'{\i} i Franqu\`es 1,
08028 Barcelona, Spain}
}

\author{Jorge N\'u\~nez}{
  address={Departament d'Astronomia i Meteorologia, Institut de Ci\`encies del
Cosmos (ICC), Universitat de Barcelona (IEEC-UB), Mart\'{\i} i Franqu\`es 1,
08028 Barcelona, Spain}
}

\begin{abstract}
Four gamma-ray binaries, namely PSR~B1259$-$63, HESS~J0632+057, HD~215227 and LS~I~+61~303, contain compact objects orbiting around massive Be stars. The nature of the compact object is only known in the case of PSR~B1259$-$63, but the other systems could also contain young non-accreting pulsars with relativistic winds. Around periastron passage the compact objects should produce significant changes in the structure of the Be discs due to gravitational forces and eventually by ram pressure from the putative pulsar wind. Indeed, variability in the $\rm{H\alpha}$ emission line has been detected in all these systems, and periodic variability in the optical photometry has been detected in two of them. However, there is lack of a systematic monitoring with accurate photometry, which could be used to constrain the shape of the disc during the periastron passage. This information is important to build accurate physical models to explain the broadband spectral energy distribution of these sources. Here we present an ongoing program to monitor the optical photometry of gamma-ray binaries and we show preliminary results for the case of HD~215227.
\end{abstract}

\maketitle


\section{Introduction}
Gamma-ray binaries are systems that comprise a young massive star and a compact object that display non-thermal emission dominated by MeV-GeV photons \citep[see][and references therein]{Moldon2012tesi, Moldon2012}. Six gamma-ray binaries are currently known. Most of these sources have been detected up to TeV energies. The spectral and brightness properties appear to be synchronised with the orbit of the binary system, and they do not show evidence of the presence of an accretion disc.

Two of the six known gamma-ray binaries, LS~5039 and 1FGL~J1018.6-5856, have an O6 massive companion. The other four binaries, LS~I~+61~303, HD~215227, HESS~J0632+057 and PSR~B1259$-$63 have a Be massive companion with a circumstellar disc in the equatorial plane.

The nature of the compact object is only known in the case of PSR~B1259$-$63, which contains a young non-accreting neutron star detected through the radio pulsations when the compact object is far away from the massive star. The study of the pulsations shows that the neutron star transforms its high rotational energy into a very powerful relativistic wind. The interaction of this wind with the wind and disc of the massive star generates non-thermal emission in the whole electromagnetic spectrum. The other five binaries probably contain young non-accreting pulsars with relativistic winds as well, although pulsations have not been detected yet \citep{Moldon2012tesi}.

In a gamma-ray binary containing a Be star the circumstellar disc is perturbed around periastron passage, due to both the gravitational forces induced by the compact object and the ram pressure of the expected relativistic pulsar wind.  \citet{Okazaki2011} and \citet{Takata2012} have studied this perturbation in the Be-pulsar system PSR~B1259$-$63 using 3D smoothed-particle hydrodynamics (SPH) simulations.

The perturbation of the Be disc in gamma-ray binaries should produce observational signatures. Indeed, all Be gamma-ray binaries show orbital variability of the $\rm{H\alpha}$ emission line close to and after the periastron passage \citep[see][and references therein]{Casares2012}. Moreover, two of these systems, namely LS~I~+61~303 and HD~215227, also display periodic variations in the optical flux, with a maximum around 0.3 orbital phases after periastron.

The main goal of this observational project is to obtain and analyse very precisely optical photometric data to model the disc properties and obtain physical parameters of the systems. Here we present preliminary results on the optical photometric monitoring of HD~215227. 

\section{Observational strategy and data reduction}
The optical observations of this project are currently made with the robotic telescope \emph{Telescope Fabra-ROA Montsec} (TFRM; Lleida, Spain) \citep{Fors2009}. The main specifications are: refurbished Baker-Nunn Camera for routine CCD robotic observations, corrector plate of 0.5~m aperture and 0.78~m primary mirror, focal ratio f/0.96, $\rm{4.4^{\circ}\times 4.4^{\circ}}$ field of view (FOV) with a pixel scale of $3.9{^\prime}{^\prime}$/pixel, colour passband filter SCHOTT GG475 (${\lambda}$ > 475 nm.), and quantum efficiency of 60\% at 550 nm.

The observational strategy for the optical monitoring of gamma-ray binaries consists on taking between 10 and 30 exposures of the order of a second (to avoid saturation) on a nightly basis (atmospheric conditions permitting).

The data reduction and analysis is made using a pipeline developed in Python following these steps: calibration of the images using {\tt PyRAF} \cite{SRScI2012}, astrometric reduction using {\tt Astrometry.net} \citep{Lang2009}, photometric extraction using {\tt SExtractor} \citep{Bertin2010}, and correction of the lightcurves using a mean differential magnitude correction method based on the principles of the {\tt BESTRED} pipeline described in \citet{Voss2006}.

The correction method is used to calibrate each frame, accounting the change in the transparency of the atmosphere as a function of the airmass, i.e., the extinction due to passing clouds, and other systematic effects. This method is resumed here:
\begin{enumerate}
\item The mean magnitude differences $\Delta m_{\rm i}$ of the preselected reference stars in all frames are computed comparing each frame to the first one: \par $\Delta m_{\rm i} = \frac{1}{n} \displaystyle \sum^n_{\rm k=1} \left( m_{\rm i,k}-m_{\rm 1,k} \right)$.
\item All the magnitudes are corrected for each frame: \par $m_{\rm i,k}^{\rm corr} = m_{\rm i,k}-\Delta m_{\rm i}$.
\item Using the corrected magnitude we select the stars with the lowest RMS between frames.
\item We compute again the correcting factor but now using only the selected stars. Finally we use this factor to calibrate the lightcurve of the gamma-ray binary.
\end{enumerate}

\section{Results on HD~215227}
\citet{Williams2010} proposed the Be star HD~215227 ($V$$\sim$8.8~mag) as the optical counterpart of the transient high-energy gamma-ray source AGL~J2241+4454 detected by {\it  AGILE} \citep{Lucarelli2010}. These authors suggested a spectral classification B3~IVne+sh and derived a distance of $2.6\pm1.0$~kpc. They also discovered an optical photometric periodicity of 60.37~d, and based on this suggested a binary nature for the source. This was later confirmed thanks to radial velocity measurements by \citet{Casares2012}.

Here we present preliminary results on the ongoing program on HD~215227, aimed to improve the results presented in \citet{Williams2010} thus allowing for a better modelling. The observations started on 12 May 2012 and here we report data until 14 September 2012, thus covering two complete 60.37d orbital cycles. During these 4 months the source could be observed in good atmospheric conditions for a total of 43 nights. 

The data reduction consisted on flat fielding and dark current subtraction, prior to the correction discussed in the previous section. The photometric precision obtained in single exposures is of 0.008 mag, which improves to 0.004 mag for nightly averages.

\begin{figure}[t!]
	\centering
        \includegraphics[width=0.84\textwidth]{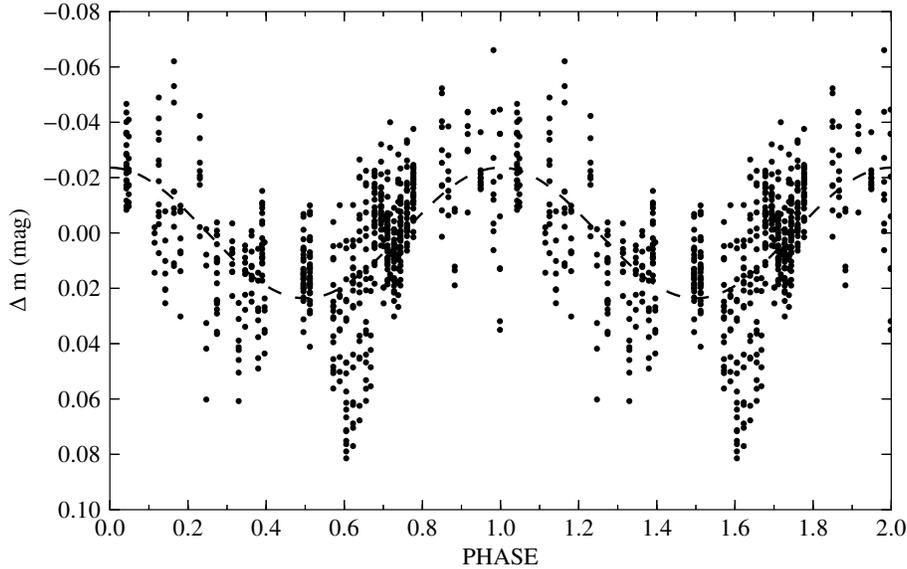}
	\caption{Lightcurve of the gamma-ray binary HD~215227 plotted as a function of cycle phase for an orbital period of 60.37~d  and phase zero at JD~2,453,243.3. The dashed line represents a sinusoidal fit to the data, with the maximum fixed at phase 0. Two cycles are displayed for clarity.}
	\label{lc}
\end{figure}

The preliminary corrected lightcurve, folded with an orbital period of 60.37 d, using $T_0=\rm{JD}~2,453,243.3$, and with an artificial offset to a mean magnitude of 0 (for comparison purposes) is shown in Fig.~\ref{lc}. The preliminary results are in good agreement with those found by \citet{Williams2010}. A sinusoidal fit with the maximum at phase 0 reveals a semiamplitude of $0.024 \pm 0.001$~mag, compatible at 2-sigma confidence level with the value of $0.020 \pm 0.002$~mag found by \citet{Williams2010}. It should be noted that the TFRM uses a colour passband filter SCHOTT~GG475, while the observations reported in \citep{Williams2010} were obtained with different interferential filters.

\section{Conclusions and outlook}
We have reported the first observational results on the ongoing program to monitor the optical photometry of gamma-ray binaries containing Be stars. We have presented preliminary results on HD~215227. The obtained photometry is accurate below 0.01~mag. We have found that our lightcurve is in good agreement with that found by \citet{Williams2010}. Further observations will alow us to better constrain the orbital period from our coherent data set, and could eventually  reveal sub-structure in the lightcurve.

We expect to obtain slightly better results in the near future, after improving the observational strategy and implementing a linearity correction that is currently under study.  We will continue the observations and the study of HD~215227. We will also carry out similar monitorings for the rest of Be gamma-ray binaries. 


\begin{theacknowledgments}
X.P.-F. and M.R. acknowledge support by the Spanish Ministerio de Econom\'{\i}a y Competitividad (MINECO) under grant FPA2010-22056-C06-02. X.P.-F. also acknowledges financial support from MINECO under grant EDU/1868/2011. M.R. also acknowledges financial support from MINECO and European Social Funds through a \emph{Ram\'on y Cajal} fellowship. O.F. acknowledges financial support from MINECO through a \emph{Juan de la Cierva} fellowship.
\end{theacknowledgments}

\bibliographystyle{aipproc}   
\bibliography{/home/gamma/garrofa/xparedes/bibtex/library.bib}

\end{document}